\documentclass[runningheads]{llncs}
\usepackage[T1]{fontenc}
\usepackage{graphicx}
\usepackage{booktabs}
\usepackage{multirow, multicol, makecell}
\usepackage{xcolor}
\usepackage{amsmath}
\usepackage{amssymb}
\usepackage{dsfont}
\usepackage[latin1]{inputenc}
\usepackage{hyperref}
\usepackage[capitalize]{cleveref}
\crefname{section}{Sec.}{Secs.}
\Crefname{section}{Section}{Sections}
\Crefname{chapter}{Chapter}{Chapters}
\Crefname{table}{Table}{Tables}
\crefname{table}{Tab.}{Tabs.}
\newcommand{\R}[1]{\textcolor[rgb]{1.00,0.00,0.00}{#1}}
\newcommand{\B}[1]{\textcolor[rgb]{0.00,0.00,1.00}{#1}}

\DeclareMathOperator*{\argmax}{arg\,max}
\DeclareMathOperator*{\argmin}{arg\,min}
\begin{document}
\title{A Study in Dataset Pruning for Image Super-Resolution}
%
%
\author{Brian B. Moser\inst{1, 2}\orcidID{0000-0002-0290-7904} \and
Federico Raue\inst{1}\orcidID{0000-0002-8604-6207} \and
Andreas Dengel\inst{1, 2}\orcidID{0000-0002-6100-8255}
}
\authorrunning{Moser et al.}
%
\institute{}
%
\maketitle              
\begin{abstract}
In image Super-Resolution (SR), relying on large datasets for training is a double-edged sword. 
While offering rich training material, they also demand substantial computational and storage resources.
In this work, we analyze dataset pruning to solve these challenges.
We introduce a novel approach that reduces a dataset to a core-set of training samples, selected based on their loss values as determined by a simple pre-trained SR model.
By focusing the training on just 50\% of the original dataset, specifically on the samples characterized by the highest loss values, we achieve results comparable to or surpassing those obtained from training on the entire dataset.
Interestingly, our analysis reveals that the top 5\% of samples with the highest loss values negatively affect the training process. 
Excluding these samples and adjusting the selection to favor easier samples further enhances training outcomes.
Our work opens new perspectives to the untapped potential of dataset pruning in image SR.
It suggests that careful selection of training data based on loss-value metrics can lead to better SR models, challenging the conventional wisdom that more data inevitably leads to better performance.

\end{abstract}
\section{Introduction}
Image Super-Resolution (SR) techniques are a cornerstone of image processing as they reconstruct High-Resolution (HR) images from their Low-Resolution (LR) counterparts \cite{moser2024diffusion,moser2023dwa,moser2023yoda,dong2015image}. 
It has wide-ranging applications, from enhancing consumer photography to improving satellite or medical imagery \cite{moser2023hitchhiker,moser2023waving,bashir2021comprehensive,valsesia2021permutation}. 
Despite its relevance, training SR models requires substantial computational resources due to large-scale datasets \cite{ganguli2022predictability,moser2023waving}. 
These datasets are pivotal for capturing the diversity of textures and patterns essential for effective upscaling, but they also pose significant storage challenges \cite{liu2021discovering}.
Recent advancements in deep learning have elevated SR techniques to new levels, with models like SwinIR and HAT setting new benchmarks for regression-based image enhancement quality \cite{liang2021swinir,chen2023activating}.

However, the success of these models often hinges on their capacity to learn from extensive and diverse training data, exacerbating the resource-intensive nature of SR model training \cite{moser2023hitchhiker,moser2023waving}.
In response to these challenges, our work explores dataset pruning as a strategy to enhance the efficiency of SR model training without compromising the quality of the output images \cite{agarwal2020contextual,paul2021deep,coleman2019selection}. 
To the best of our knowledge, efforts to apply dataset pruning to image SR tasks have been scarce, except for the notable contribution made by \textit{Ding et al.} \cite{ding2023not}, which we will discuss and improve on in this work.
The concept of dataset pruning involves reducing the size of the training dataset by selectively identifying a subset of samples that are most informative for the optimization process. 
This approach is promising for mitigating the storage burden of large datasets while preserving or improving the training performance of SR models \cite{ganguli2022predictability,moser2022less}.

Our contribution is twofold. 
First, we propose a novel loss-value-based sampling method for dataset pruning in image SR, leveraging a simple pre-trained SR model, namely SRCNN \cite{dong2015image}.
Our method contrasts traditional approaches that indiscriminately use the entirety of available data, i.e., DIV2K \cite{agustsson2017ntire}.
Secondly, we empirically demonstrate that training SR models on a pruned dataset - comprising 50\% of the original dataset selected based on their loss values - can achieve comparable or superior performance to training on the full dataset. 
Refining this selection by excluding the top 5\% hardest samples, which we found were counterproductive, further enhances model training efficiency.

Through this work, we aim to spark a paradigm shift in how training datasets are curated for SR tasks. 
We advocate for a loss-value-driven approach to dataset pruning. 
Our strategy significantly reduces the storage requirements of SR model training and offers a scalable solution that can adapt to the evolving complexities and requirements of image SR.

\section{Related Work}

Dataset pruning is particularly interesting to deep learning \cite{agarwal2020contextual,paul2021deep,coleman2019selection}. 
It focuses on training set size reduction while attempting to maintain or even enhance the performance of models. 
This process is not just about economizing on computational storage; it also aims to improve model generalization by eliminating redundant or less informative samples \cite{katharopoulos2018not,hinton2015distilling}. 
Various methodologies have explored this concept, including importance sampling, core-set selection, and data distillation. 
In the following, we explore key contributions to dataset pruning and provide insights into its development.

\textbf{Importance Sampling.} One key approach in dataset pruning is importance sampling, where the idea is to prioritize training on samples deemed more important for the model. 
Works such as \textit{Katharopoulos et al.} \cite{katharopoulos2018not} have explored adaptive sampling methods that dynamically adjust the probability of selecting each sample based on the model's current state. 
These methods aim to focus computational effort where it is most needed.
However, by weighting samples, importance sampling does not reduce the training set size as we do.

\textbf{Data distillation.} Data distillation is a technique that generates a condensed and synthetic version of the training data, often through knowledge distillation \cite{hinton2015distilling}, where a smaller dataset is created to capture the essence of the original data. The work by \textit{Wang et al.} \cite{wang2018dataset} on dataset distillation demonstrates how training on a distilled dataset can achieve comparable performance to training on the entire dataset, significantly reducing the computational burden regarding convergence and storage. 
Since then, various exciting developments have been made in dataset distillation \cite{moser2024latent,cazenavette2023generalizing}.
Nevertheless, existing approaches to dataset distillation are predominantly aimed at capturing critical semantic details for image classification purposes \cite{zhao2022synthesizing,cazenavette2022dataset,nguyen2020dataset,zhao2020dataset}. 
Given the significant differences between image classification and image SR, applying dataset distillation techniques directly to SR tasks presents a considerable challenge \cite{liu2021discovering}.

\textbf{Core-Set Selection.} This approach involves identifying a subset of the training data that is a good representation of the entire dataset. 
Therefore, it reduces the dataset size in contrast to importance sampling. 
\textit{Sener et al.} \cite{sener2017active} introduced an optimization framework that selects samples constituting a core-set for training deep neural networks. 
This method minimizes the maximum loss over the dataset, ensuring the selected subset is as informative as the original dataset.
Since then, core-set selection, also called proxy datasets, has been further developed in various fields, such as image classification or neural architecture search \cite{moser2022less,ding2023not,coleman2019selection,shleifer2019using}. 

To the best of our knowledge, image SR tasks have primarily remained untouched by initiatives in dataset pruning, with a notable exception of the work conducted by \textit{Ding et al.} \cite{ding2023not}.
The authors suggest using the Sobel filter to reduce the dataset, focusing specifically on selecting samples with rich textures. 
They further refine their selection by clustering these texture-rich samples to ensure a variety of textures is represented. 
In contrast, we opt for sampling based on SR reconstruction loss values. 
In the experiments section, we will demonstrate that our strategy, which prioritizes samples according to their SR reconstruction loss, performs better than a method based on Sobel filter selection.

\section{Methodology}

This section introduces our method for optimizing SR model training using dataset pruning.
Our approach is found on the premise that not all samples in a dataset contribute equally to the learning process in the context of SR tasks \cite{sener2017active,katharopoulos2018not,moser2022less}. 
By carefully selecting a core-set of samples that are most informative for SR model training, we aim to enhance the learning process.

After introducing the concept of core-sets and how they are sampled, we present our loss-based sampling method, which leverages a pre-trained SR model - a simple SRCNN \cite{dong2015image} - to estimate the complexity in reconstructing HR samples from their LR counterparts. 
By focusing on challenging samples for the SR model, as indicated by higher loss values, we hypothesize that the model can learn more effectively, thereby improving its performance on unseen data.

\subsection{Core-Sets}
Consider a dataset $\mathcal{D} = \{ \left( \mathbf{x}_i, \mathbf{y}_i \right) \}$ with a total number of elements denoted as $\text{N}_\mathcal{D}$. 
In this context, $\mathbf{x}_i$, where $0 \leq i < \text{N}_\mathcal{D}$, represents the $i$-th LR sample and $\mathbf{y}_i$ its corresponding HR counterpart.
We define a subset of $\mathcal{D}$, denoted by $\mathcal{D}_r \subset \mathcal{D}$, as a core-set with a size of $\text{N}_{\mathcal{D}_r}$. 
The proportion $r \in \left( 0, 1 \right)$ specifies the size of $\mathcal{D}_r$ relative to $\mathcal{D}$: $\text{N}_{\mathcal{D}_r} \approx r \cdot \text{N}_\mathcal{D}$. 
This approximation accounts for instances where the dataset size cannot be divided evenly.
The core-set acts between the original dataset and the SR model to enhance the SR output quality by focusing on important samples during training.
The objective is to construct a core-set $\mathcal{D}_r \subset \mathcal{D}$ such that its size $\text{N}_{\mathcal{D}_r} \approx r \cdot \text{N}_\mathcal{D}$, with $r \in \left( 0, 1 \right)$.
To achieve this for any chosen $r$, the sampling strategy must satisfy the condition

\begin{equation}
    \text{N}_{\mathcal{D}_r} = \sum_{ \left(\mathbf{x}_i, \mathbf{y}_i \right) \in \mathcal{D}} \mathds{1}_{\mathcal{D}_r} \left( \mathbf{x}_i \right) \approx r \cdot \text{N}_\mathcal{D},
    \label{eq:sampling}
\end{equation}

\noindent
where $\mathds{1}_{\mathcal{D}_r}: \mathcal{D} \rightarrow \{ 0, 1\}$ is the indicator that determines whether an element belongs to the core-set $\mathcal{D}_r$ within the larger dataset $\mathcal{D}$.

The concrete realization of the core-set selection depends on the sampling mechanism. \textit{Ding et al.} \cite{ding2023not} proposed to use the Sobel filter to identify texture-rich training samples. In contrast, we argue that a loss-value-based sampling is more efficient, which we will explain next and demonstrate empirically in the experiments section.

\begin{figure}[!t]
  \centering
  \includegraphics[width=\linewidth]{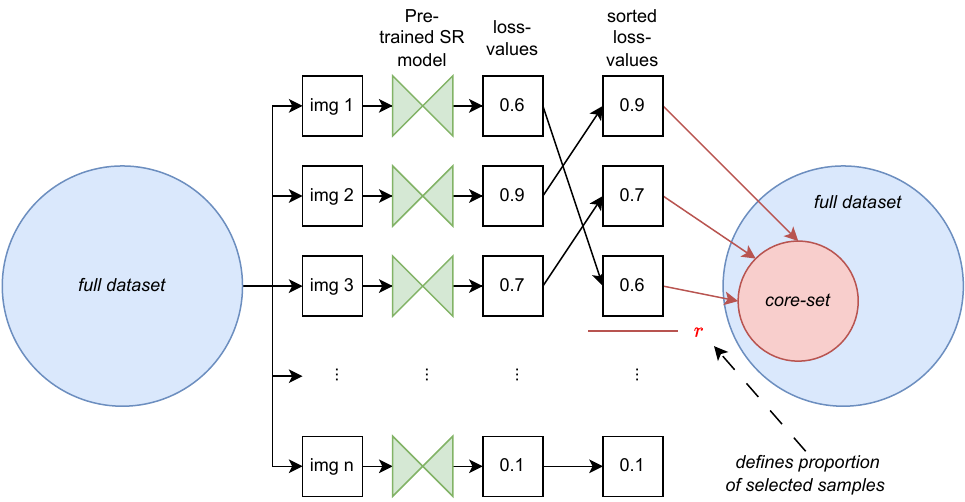}
  \caption{\label{fig:concept}Illustration of our loss-value-based core-set selection for image SR. Initially, the full dataset undergoes evaluation through a pre-trained SR model to calculate loss values for each image pair. These loss values are then sorted to identify samples with varying degrees of reconstruction difficulty. A pre-defined proportion $r$ of these samples is selected to form a core-set.}
\end{figure}

\subsection{Loss-Value-based Sampling}
In image SR, one approximates $\mathbf{y}_i \approx \mathcal{M} \left( d \left( \mathbf{y}_i \right) \right)$ with a SR model $\mathcal{M}$ and a degradation function $d$ that represents the relationship between the HR and LR space (i.e., $d \left( \mathbf{y}_i \right) = \mathbf{x}_i$).
As a result, a trained SR model can provide a distance metric based on a loss function.
Let $\mathcal{L}: \mathbb{R}^{h \times w \times c} \times \mathbb{R}^{h\times w \times c} \rightarrow \mathbb{R}$ be a loss function, e.g., Mean Squared Error (MSE), with $h$, $w$, $c$ as the height, width, and channel size, respectively.
Given $r \in \left( 0, 1\right)$, we can derive a core-set with
\begin{equation}
    \label{eq:easy}
    \mathcal{D}^{\text{ASC}}_r = \argmin_{\substack{\mathcal{D}' \subset \mathcal{D}, \\ \text{s.t. }\mid \mathcal{D}'\mid \approx r \cdot \text{N}_\mathcal{D}}} \sum_{\left(\cdot, \mathbf{y}_i \right) \in \mathcal{D}'} \mathcal{L} \left( \mathcal{M} \left( d \left( \mathbf{y}_i \right)\right), \mathbf{y}_i\right),
\end{equation}
\noindent
where samples with high loss values are removed, denoted as ascending sampling (ASC).
Likewise, we can define a sampling method based on removing the lowest loss values, thereby concentrating on hard samples by 
\begin{equation}
    \label{eq:hard}
    \mathcal{D}^{\text{DES}}_r = \argmax_{\substack{\mathcal{D}' \subset \mathcal{D}, \\ \text{s.t. }\mid \mathcal{D}'\mid \approx r \cdot \text{N}_\mathcal{D}}} \sum_{\left(\cdot, \mathbf{y}_i \right) \in \mathcal{D}'} \mathcal{L} \left( \mathcal{M} \left( d \left( \mathbf{y}_i \right)\right), \mathbf{y}_i\right),
\end{equation}
denoted as descending sampling (DES). The concept is illustrated in \autoref{fig:concept}.
In the experiments section, we will determine whether ascending or descending sampling is preferable and which $r$ value is beneficial.
Intuitively, ascending sampling includes less complex, monochromatic training samples first, whereas descending sampling favors texture-rich, multi-colored training samples.
In the following, we will use a simple pre-trained SRCNN \cite{dong2015image} (composed of three layers of convolutions) and MSE for the loss calculation in \autoref{eq:easy} and \autoref{eq:hard}.

\section{Experiments}
In this section, we empirically evaluate the performance of our loss-value-based sampling. 
We start by benchmarking whether ascending (ASC) or descending (DES) sampling is more beneficial.
We also evaluate whether maintaining the same number of training steps is critical.
Next, we compare our loss-value-based sampling with the sampling mechanism exchanged by using Sobel filters instead, as suggested by \textit{Ding et al.} \cite{ding2023not}.
Finally, we analyze the pruned dataset and suggest a refined version. 
We will evaluate the original and refined core-set with state-of-the-art datasets and methods. 

\subsection{Datasets}
Our method is assessed using well-established SR datasets. 
The DIV2K dataset \cite{agustsson2017ntire} served as our primary source for training data, from which we extracted sub-images according to standard practice in literature \cite{bashir2021comprehensive,moser2024diffusion}. 
As a result, we derive around 32K HR training samples from 800 2K HR images.
LR samples are computed by following the standard procedure using bicubic interpolation and anti-aliasing \cite{MATLAB:2017b,moser2023hitchhiker}.
These sub-images formed the basis for selecting the core-sets. 
For the evaluation phase, we utilized the test datasets Set5 \cite{Set5}, Set14 \cite{Set14}, BSDS100 \cite{BSD100}, and Urban100 \cite{Urban100}.
We assess our experiments based on two metrics: Peak Signal-to-Noise Ratio (PSNR) and Structural Similarity Index (SSIM), where higher values represent better image quality.

\subsection{Results}

\begin{table}[!htp]

    \caption{\label{tab:ablTab}
    Comparison of ascending and descending sampling on BSD100 with $2\times$ scaling. Values with at least the same or better performance than their corresponding performance on the full dataset are highlighted in \R{red}.
    }
    \centering
    \resizebox{\textwidth}{!}{%
    \begin{tabular}{|c|c | c c | c c | c c | c c | c c|}
    \hline
    \multirow{ 2}{*}{\makecell{Method}} & \multirow{ 2}{*}{\makecell{Train\\Steps}} & \multicolumn{2}{c|}{{FSRCNN} \cite{dong2016accelerating}} & \multicolumn{2}{c|}{{DRRN} \cite{tai2017image}} & \multicolumn{2}{c|}{{IDN} \cite{hui2018fast}} & \multicolumn{2}{c|}{{RDN} \cite{zhang2018residual}} & \multicolumn{2}{c|}{{SwinIR} \cite{liang2021swinir}}\\
    \cline{3-12}
    & & {PSNR}  & {SSIM}  & {PSNR}  & {SSIM}  & {PSNR}  & {SSIM}  & {PSNR}  & {SSIM}  & {PSNR}  & {SSIM} \\
    \hline
    full & 15,608 & 31.09 & 0.8955 & 29.42 & 0.8781 & 31.98 & 0.9076 & 32.25 & 0.9103 & 32.23 & 0.9103\\
    \hline
    75\% ASC & 11,706 & 30.84 & 0.8928 & 28.11 & 0.8725 & 31.90 & 0.9065 & 32.21 & 0.9101 & 32.13 & 0.9091\\
    50\% ASC & 7,804 & 30.25 & 0.8862 & 22.28 & 0.8370 & 31.67 & 0.9040 & 32.02 & 0.9080 & 31.90 & 0.9067\\
    25\% ASC & 3,902 & 28.38 & 0.8303 & 14.66 & 0.7293 & 31.23 & 0.8976 & 31.66 & 0.9039 & 30.97 & 0.8957\\
    \hline
    75\% DES & 11,706 & \R{31.11} & \R{0.8960} & 29.32 & \R{0.8783} & 31.96 & 0.9071 & \R{32.26} & \R{0.9107} & \R{32.24} & \R{0.9104}\\
    50\% DES & 7,804 & \R{31.20} & \R{0.8968} & 29.40 & \R{0.8795} & 31.94 & 0.9070 & \R{32.25} & \R{0.9105} & 32.21 & 0.9101\\
    25\% DES & 3,902 & 30.98 & 0.8937 & 29.32 & \R{0.8791} & 31.81 & 0.9052 & 32.22 & 0.9099 & 32.13 & 0.9088\\
    \hline
    25\% DES & 7,804 & \R{31.11} & \R{0.8969} & 29.39 & \R{0.8804} & 31.94 & 0.9068 & 32.24 & \R{0.9104} & 32.20 & 0.9096\\   
    25\% DES & 15,608 & \R{31.35} & \R{0.8986} & \R{29.46} & \R{0.8810} & 31.98 & \R{0.9076} & \R{32.26} & \R{0.9104} & \R{32.24} & 0.9101\\ 
    \hline
    50\% DES & 15,608 & \R{31.30} & \R{0.8983} & 29.39 & \R{0.8795} & \R{32.00} & \R{0.9076} & \R{32.28} & \R{0.9108} & \R{32.27} & \R{0.9104}\\ 
    \hline
    \end{tabular}}
\end{table}

\subsubsection{Ascending versus Descending Sampling.}

We compare ascending (ASC) and descending (DES) sampling methods on the BSD100 dataset with $2\times$ scaling in \autoref{tab:ablTab}. 
Our primary focus is understanding their impact on various image SR models, namely FSRCNN \cite{dong2016accelerating}, DRRN \cite{tai2017image}, IDN \cite{hui2018fast}, RDN \cite{zhang2018residual}, and SwinIR \cite{liang2021swinir}. 
The various models evaluate different categories of SR models, namely simple CNN, recursive CNN, residual CNNs, and Transformer-based \cite{moser2023hitchhiker}. 

We observe a general trend of declining performance, i.e., worse image enhancement quality, across all models when ascending sampling (ASC) is applied at reduced dataset sizes. 
Specifically, at 25\% dataset size, the performance drop becomes more pronounced, with the PSNR dropping from 31.09 to 28.38 for FSRCNN and from 29.42 to 14.66 for DRRN.
This indicates the ineffectiveness of ascending sampling in preserving model performance.

\begin{figure}[!t]
  \centering
  \includegraphics[width=\linewidth]{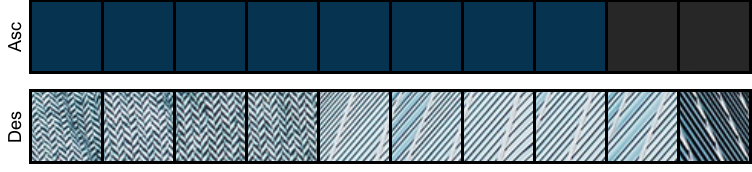}
  \caption{\label{fig:effsatops}Comparison of top-selected samples by ascending and descending sampling. We can observe that descending sampling selects primarily training patches with high textural details, whereas ascending sampling focuses on monochromatic samples.}
\end{figure}

In contrast, descending sampling generally outperforms or matches the full dataset performance, especially notable in the 75\% and 50\% dataset sizes. 
For instance, FSRCNN shows improved results at 75\% DES and even higher scores at 50\% DES, surpassing the full dataset baseline. 

Note that decreased data size leads to fewer training steps if the number of epochs is fixed.
Consequently, with fixed epochs, i.e., 200, SR models train effectively shorter because a single epoch has fewer training iterations due to the reduced data size. 
Remarkably, applying more epochs to match the number of training steps on the original dataset further enhances the results.
Using 25\% DES with 4x epochs, which is equal to the number of training steps employed in the full dataset, achieves the highest PSNR and SSIM for several models. 

Our findings highlight that hard samples, defined by high loss, are crucial for maintaining or enhancing SR model performance. 
Therefore, texture-rich samples, contrary to monochromatic easy samples (see \autoref{fig:effsatops}), are significant for training a SR model effectively.
Additionally, maintaining the number of training iterations is essential, rather than merely focusing on reducing the dataset size for reduced training time. 
By keeping training iterations consistent, descending sampling effectively leverages a reduced but more potent subset of the original dataset, leading to improved or comparable performance across all evaluated SR models.
Our findings about keeping the same training iterations present a distinct divergence from those reported by \textit{Ding et al.} \cite{ding2023not}. 
We speculate that this discrepancy may arise from using fewer training epochs, which likely prevented their models from achieving full convergence.
Moving forward, we will further explore our loss-based sampling approach by maintaining a consistent number of iterations and employing the descending sampling technique.

\begin{table*}[t]
\centering
\begin{center}
\caption{Quantitative comparison between sampling based on loss-value and Sobel filter (average PSNR/SSIM) with SwinIR for classical image SR on benchmark datasets ($2\times$ scaling, 50\% pruning). 
The best performance is highlighted in \R{red}.}%

\label{tab:sobel_vs_loss}
\begin{tabular}{|c|cc|cc|cc|cc|cc|}
\hline
\multirow{2}{*}{\makecell{Sampling\\Method}} &  \multicolumn{2}{c|}{Set5~\cite{Set5}} &  \multicolumn{2}{c|}{Set14~\cite{Set14}} &  \multicolumn{2}{c|}{BSD100~\cite{BSD100}} &  \multicolumn{2}{c|}{Urban100~\cite{Urban100}} &  \multicolumn{2}{c|}{Manga109~\cite{matsui2017sketch}} 
\\
\cline{2-11}
& PSNR & SSIM & PSNR & SSIM & PSNR & SSIM & PSNR & SSIM & PSNR & SSIM 
\\
\hline
\hline
loss-based
& \R{38.32}
& \R{0.9619}
& \R{34.15}
& \R{0.9232}
& \R{32.45}
& \R{0.9038}
& \R{33.43}
& \R{0.9396}
& \R{39.55}
& \R{0.9790}
\\
sobel-based
& {38.26}
& {0.9612}
& {34.09}
& {0.9225}
& {32.44}
& {0.9037}
& {33.38}
& {0.9392}
& {39.54}
& {0.9789}
\\
\hline             
\end{tabular}
\end{center}
\vspace{1mm}
\end{table*}

\subsubsection{Loss-based versus Sobel-based}
As introduced by \textit{Ding et al.} \cite{ding2023not}, utilizing the Sobel filter represents another strategy for curating a core-set for image SR. 
Therefore, we compare our loss-based sampling method with the Sobel filter approach.
From \autoref{tab:sobel_vs_loss}, it is evident that the loss-based sampling method consistently outperforms the Sobel filter-based approach across all datasets. 
Specifically, for the Set5 dataset, loss-based sampling achieves a PSNR of 38.32 and an SSIM of 0.9619, compared to 38.26 (PSNR) and 0.9612 (SSIM) for the Sobel-based method. 
This trend continues across the Set14, BSD100, and Urban100 datasets, where loss-based sampling performs superiorly in both PSNR and SSIM metrics.
By focusing on loss values, we prioritize samples the model finds challenging to reconstruct, potentially leading to a more robust learning process and improved SR performance. 
In contrast, the Sobel filter-based approach, which selects samples based on texture richness, might overlook other crucial aspects contributing to the overall quality and effectiveness of SR reconstruction.

\begin{figure}[!t]
  \centering
  \includegraphics[width=.7\linewidth]{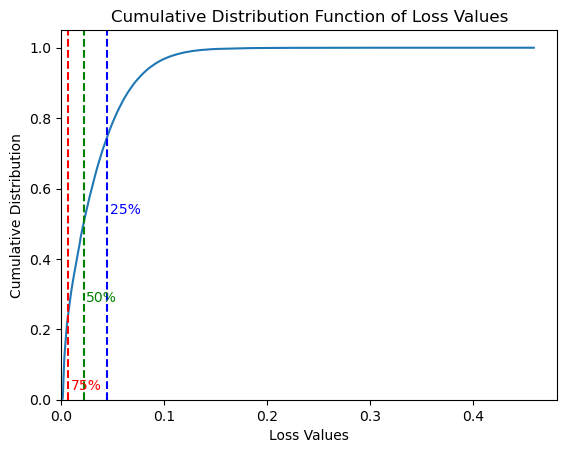}
  \caption{\label{fig:cdf}Cumulative Loss-Value Distribution (sorted). Vertical lines represent different descending sampling endpoints for 25\%, 50\%, and 75\% sampling. The right side of the respective vertical line shows the loss values included and found within the corresponding core-sets.}
\end{figure}

\subsubsection{Analyzing Loss Values.}
This section examines the core-sets derived by descending sampling more closely. 
More specifically, we derive the loss values found by applying descending sampling at 25\%, 50\%, and 75\%.
\autoref{fig:cdf} shows the result.  
The distribution demonstrates that certain samples, characterized by notably higher loss values, are consistently included across all core-sets (see the long tail on the right side of the vertical lines). 
We theorize that these samples, possibly due to their high noise levels, could be detrimental to the training of SR models. 
To address this, we suggest refining the initially derived core-set by adjusting the selection threshold by 5\% to favor samples with lower loss values and exclude those with the highest losses.

\begin{figure}[!t]
  \centering
  \includegraphics[width=\linewidth]{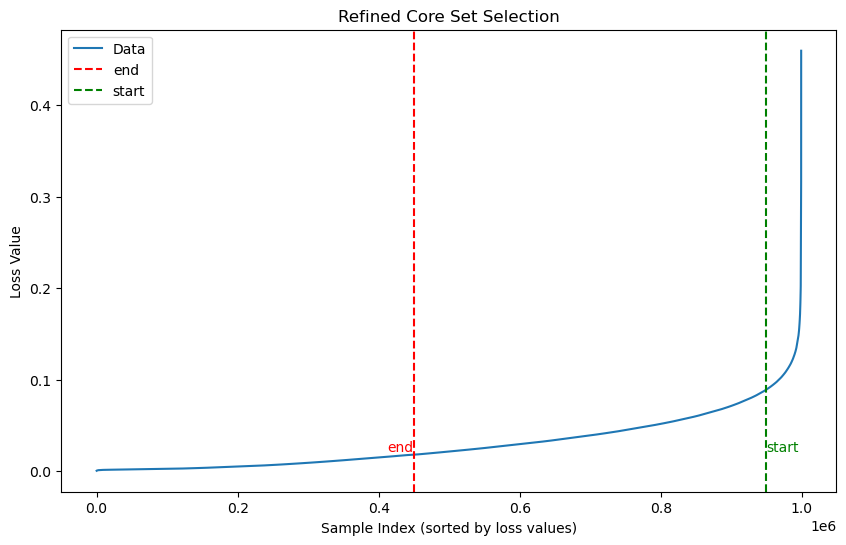}
  \caption{\label{fig:refined}Refined Core Set Proposal. This strategy selects the top 50\% most challenging samples but modifies the selection by shifting the inclusion threshold by 5 \% towards samples with lower loss values, aiming for a more balanced core-set. In other words, we keep 50 \% of the hardest samples in our core-set after excluding the top 5 \% from the dataset.}
\end{figure}

The modified core-set and associated loss values within this refined set are depicted in \autoref{fig:refined}; see the area between the vertical lines denoted by start and end. 
This approach involves retaining 50\% of the samples initially considered the most challenging but adjusting the selection to slightly favor easier samples (those with lower loss values) by shifting the inclusion criteria by 5\%.
In other words, the core-set still contains 50\% of the original dataset's hardest samples, except the top 5\% hardest.
We will evaluate both core-sets, the original and the refined, in the following with state-of-the-art benchmarks.

\begin{table*}[!t]
\centering
\begin{center}
\caption{Quantitative comparison (average PSNR/SSIM) with state-of-the-art methods for classical image SR on benchmark datasets (Set5, Set14, BSDS100, and Urban100). 
These experiments were evaluated under the lens of various scaling factors, specifically 2, 3, and 4.
We trained SwinIR on our pruned datasets with 50 \% of the size of the original datasets.
Moreover, we included our refined version of our core-set, which excludes the top 5\% hardest samples (see ref. 50 \%).
The best and second best performances are in \R{red} and \B{blue} colors, respectively.
As a result, training SwinIR on half of the original datasets leads to comparable and, in most cases, superior performance.}

\label{tab:sr_results}
\begin{tabular}{|l|c|c|cc|cc|cc|cc|}
\hline
\multirow{2}{*}{Method} & \multirow{2}{*}{Scale} & \multirow{2}{*}{\makecell{Train\\Set}} &  \multicolumn{2}{c|}{Set5~\cite{Set5}} &  \multicolumn{2}{c|}{Set14~\cite{Set14}} &  \multicolumn{2}{c|}{BSD100~\cite{BSD100}} &  \multicolumn{2}{c|}{Urban100~\cite{Urban100}} 
\\
\cline{4-11}
&  &  & PSNR & SSIM & PSNR & SSIM & PSNR & SSIM & PSNR & SSIM 
\\
\hline
\hline
RCAN~\cite{zhang2018image} & $\times$2 & full %
& {38.27}
& {0.9614}
& {34.12}
& {0.9216}
& {32.41}
& {0.9027}
& {33.34}
& {0.9384}
\\  
SAN~\cite{dai2019SAN} & $\times$2 & full %
& {38.31}
& {0.9620}
& {34.07}
& {0.9213}
& {32.42}
& {0.9028}
& {33.10}
& {0.9370}
\\
IGNN~\cite{zhou2020IGNN} & $\times$2 & full %
& {38.24}
& {0.9613}
& {34.07}
& {0.9217}
& {32.41}
& {0.9025}
& {33.23}
& {0.9383}
\\
HAN~\cite{niu2020HAN} & $\times$2 & full %
& {38.27}
& {0.9614}
& {34.16}
& {0.9217}
& {32.41}
& {0.9027}
& {33.35}
& {0.9385}
\\ 
NLSA~\cite{mei2021NLSA} & $\times$2 & full %
& \B{38.34} 
& {0.9618} 
& 34.08 
& {0.9231}
& 32.43 
& 0.9027 
& {33.42}
& {0.9394}
\\
\textbf{SwinIR}~\cite{liang2021swinir}  & $\times$2  & full
& \R{38.35}
& \R{0.9620}
& {34.14}
& {0.9227}
& {32.44}
& {0.9030}
& {33.40}
& {0.9393}
\\
\hline 
\textbf{SwinIR}~\cite{liang2021swinir}  & $\times$2  & 50\%
& \R{38.35}
& \R{0.9620}
& \B{34.20}
& \B{0.9230}
& \B{32.46}
& \B{0.9039}
& \B{33.47}
& \B{0.9399}
\\
\textbf{SwinIR}~\cite{liang2021swinir}  & $\times$2  & ref. 50\%
& \B{38.34}
& \B{0.9619}
& \R{34.23}
& \R{0.9236}
& \R{32.48}
& \R{0.9041}
& \R{33.52}
& \R{0.9401}
\\
\hline                 
\hline
RCAN~\cite{zhang2018image}& $\times$3   & full
& {34.74}
&{0.9299}
& {30.65}
& {0.8482}
& {29.32}
& {0.8111}
& {29.09}
& {0.8702}
\\
SAN~\cite{dai2019SAN} & $\times$3   & full
& {34.75}
& {0.9300}
& {30.59}
& {0.8476}
& {29.33}
& {0.8112}
& {28.93}
& {0.8671}
\\
IGNN~\cite{zhou2020IGNN} & $\times$3  & full
& {34.72}
& {0.9298}
& {30.66}
& {0.8484}
& {29.31}
& {0.8105}
& {29.03}
& {0.8696}
\\
HAN~\cite{niu2020HAN}  & $\times$3   & full
& {34.75}
& {0.9299}
& {30.67}
& {0.8483}
& {29.32}
& {0.8110}
& {29.10}
& {0.8705}
\\
NLSA~\cite{mei2021NLSA} & $\times$3  & full
& 34.85 
& 0.9306 
& 30.70 
& 0.8485 
& 29.34 
& 0.8117 
& {29.25}
& {0.8726}
\\
\textbf{SwinIR}~\cite{liang2021swinir}  & $\times$3  & full
& \R{34.89}
& \R{0.9312}
& \R{30.77}
& \B{0.8503}
& \R{29.37}
& {0.8124}
& \R{29.29}
& \R{0.8744}
\\
\hline 
\textbf{SwinIR}~\cite{liang2021swinir}  & $\times$3  & 50\%
& \B{34.87}
& \B{0.9307}
& {30.70}
& \B{0.8503}
& \B{29.35}
& \B{0.8134}
& {29.19}
& {0.8731}
\\
\textbf{SwinIR}~\cite{liang2021swinir}  & $\times$3  & ref. 50\%
& {34.84}
& {0.9306}
& \B{30.75}
& \R{0.8506}
& \R{29.37}
& \R{0.8137}
& \B{29.25}
& \B{0.8740}
\\
\hline
\hline
RCAN~\cite{zhang2018image}& $\times$4  & full
& {32.63}
& {0.9002}
& {28.87}
&{0.7889}
& {27.77}
& {0.7436}
&{26.82}
& {0.8087}
\\ 
SAN~\cite{dai2019SAN} & $\times$4  & full
& {32.64}
& {0.9003}
& {28.92}
& {0.7888}
& {27.78}
& {0.7436}
& {26.79}
& {0.8068}
\\
IGNN~\cite{zhou2020IGNN}  & $\times$4  & full
& {32.57}
& {0.8998}
& {28.85}
& {0.7891}
& {27.77}
& {0.7434}
& {26.84}
& {0.8090}
\\
HAN~\cite{niu2020HAN}  & $\times$4  & full
& {32.64}
& {0.9002}
& {28.90}
& {0.7890}
& {27.80}
& {0.7442}
& {26.85}
& {0.8094}
\\
NLSA~\cite{mei2021NLSA} & $\times$4 & full
& 32.59 
& 0.9000 
& 28.87 
& 0.7891 
& 27.78 
& 0.7444 
& {26.96}
& {0.8109}
\\
\textbf{SwinIR}~\cite{liang2021swinir}  & $\times$4  & full
& \B{32.72}
& \R{0.9021}
& \R{28.94}
& \R{0.7914}
& \R{27.83}
& {0.7459}
& \R{27.07}
& \R{0.8164}
\\
\hline 
\textbf{SwinIR}~\cite{liang2021swinir}  & $\times$4  & 50\%
& {32.71}
& \B{0.9013}
& \R{28.91}
& \B{0.7908}
& {27.80}
& \B{0.7466}
& {26.91}
& \B{0.8113}
\\
\textbf{SwinIR}~\cite{liang2021swinir}  & $\times$4  & ref. 50\%
& \R{32.75}
& {0.9012}
& {28.87}
& {0.7903}
& \B{27.81}
& \R{0.7469}
& \B{26.92}
& {0.8111}
\\
\hline             
\end{tabular}
\end{center}
\vspace{1mm}
\end{table*}

\subsubsection{Benchmark with State-of-the-Art.}

We evaluate the quality of our core-sets by training the state-of-the-art SR model SwinIR \cite{liang2021swinir} on them. 
The evaluation includes other state-of-the-art methods for classical image SR across benchmark datasets, including Set5~\cite{Set5}, Set14~\cite{Set14}, BSD100~\cite{BSD100}, and Urban100~\cite{Urban100}, with scaling factors of $2\times$, $3\times$, and $4\times$.
\autoref{tab:sr_results} reports the results with notable performances being color-coded, with the best and second-best results marked in red and blue, respectively.

Interestingly, when SwinIR is trained on 50\% of the dataset for $2\times$ scaling, it maintains its superior performance on Set5 and improves upon the full dataset results on Set14, BSD100, and Urban100, suggesting efficient learning from a pruned dataset. 
The refined 50\% core-set further improves performance, pushing the boundaries on Set14, BSD100, and Urban100 to achieve the highest metrics.
Again, This indicates that a pruned dataset can match or even surpass full dataset training outcomes.

For a $3\times$ scaling factor, the core-sets demonstrate competitive or superior performance compared to the full dataset, especially highlighted in the Set14 and BSD100 datasets. 
This reinforces the effectiveness of dataset pruning in enhancing model efficiency without significantly compromising output quality.

The core-sets yield closely competitive results at the most challenging scaling factor of 4.
The refined core-set offers the best PSNR on BSD100 and closely matches the full dataset's performance on Urban100.
A general observation is that the refined core-set outperforms the original 50\% core-set in most cases.

These findings underscore the potential of dataset pruning strategies, especially when applied to sophisticated models like SwinIR. 
The experiments suggest that thoughtful pruning can achieve comparable or superior performance using only a fraction of the training data. 

\section{Conclusion and Future Work}
By introducing and comparing loss-value-based sampling strategies, our study highlights the potential of dataset pruning to maintain and, in most instances, enhance the performance of SR models while substantially reducing the computational storage. 
We use a simple pre-trained SR model, SRCNN, to determine which samples to remove based on a mean squared error loss.
Our findings, particularly with the descending sampling method, underscore the value of selectively curating training datasets to include samples challenging for the model during training.
In other words, as defined by loss values, training SR models on hard samples is more beneficial than training on easy samples.
Furthermore, we concluded that loss-value-based sampling performed better than Sobel filter-based sampling.
Moreover, we showed that our refined core-set, which excludes the top 5 \% of hardest samples, further improves the performance.
We have validated our approach against several benchmarks using current state-of-the-art models.
Our experiments also verify our hypothesis on several SR models, including leading models like SwinIR, and on several datasets.

For future work, we see significant potential in advancing our dataset pruning strategies by integrating more nuanced measures of sample difficulty. This could involve leveraging insights from model uncertainty or incorporating adaptive feedback loops during training to adjust the core-set dynamically. Such refinements could pave the way for more efficient training methods, optimizing computational resources while achieving superior SR model performance.

\section*{Acknowledgements} 
This work was supported by the BMBF project SustainML (Grant 101070408).

\bibliographystyle{splncs04}
\bibliography{mybib}
\end{document}